\documentclass[journal]{IEEEtran}
\usepackage{amsmath}
\usepackage{amssymb}
\usepackage{amsthm}
\usepackage{cite}
\usepackage{color}
\usepackage{xcolor}
\usepackage{caption}
\usepackage{epsfig,latexsym}
\usepackage{float}
\usepackage{fancyhdr}
\usepackage{graphicx}
\usepackage{indentfirst}
\usepackage{mdwmath}
\usepackage{mdwtab}
\usepackage{subfigure}
\usepackage{setspace}
\usepackage{times}
\usepackage{url}
\usepackage{multirow}

\newtheorem{remark}{Remark}
\newtheorem{theorem}{Theorem}

\newtheorem{lemma}{Lemma}

\newtheorem{corollary}{Corollary}

\begin{document}

\title{NOMA Enhanced Terrestrial and Aerial IoT Networks with Partial CSI}
\author{Tianwei Hou,~\IEEEmembership{Student Member,~IEEE,}
        Yuanwei Liu,~\IEEEmembership{Senior Member,~IEEE,} \\
        Zhengyu Song,
        Xin Sun,
        and Yue Chen,~\IEEEmembership{Senior Member,~IEEE,}

\thanks{This work is supported by Beijing Natural Science Foundation under Grant 4194087. Corresponding author: Zhengyu Song.}
\thanks{T. Hou, Z. Song and X. Sun are with the School of Electronic and Information Engineering, Beijing Jiaotong University, Beijing 100044, China (email: 16111019@bjtu.edu.cn, songzy@bjtu.edu.cn, xsun@bjtu.edu.cn).}
\thanks{Y. Liu and Yue Chen are with School of Electronic Engineering and Computer Science, Queen Mary University of London, London E1 4NS, U.K. (e-mail: yuanwei.liu@qmul.ac.uk, yue.chen@qmul.ac.uk).}
}

\maketitle

\begin{abstract}
This article investigates a non-orthogonal multiple access (NOMA) enhanced Internet of Things (IoT) network. In order to provide connectivity, a novel cluster strategy is proposed, where multiple devices can be served simultaneously. Two potential scenarios are investigated: 1) NOMA enhanced terrestrial IoT networks and 2) NOMA enhanced aerial IoT networks. We utilize stochastic geometry tools to model the spatial randomness of both terrestrial and aerial devices. New channel statistics are derived for both terrestrial and aerial devices. The exact and the asymptotic expressions in terms of coverage probability are derived. In order to obtain further engineering insights, short-packet communication scenarios are investigated. From our analysis, we show that the performance of NOMA enhanced IoT networks is capable of outperforming OMA enhanced IoT networks. Moreover, based on simulation results, there exists an optimal value of the transmit power that maximizes the coverage probability.
\end{abstract}


\section{Introduction}

In recent years, a growing number of Internet of Things (IoT) devices are being connected to the internet at an unprecedented rate~\cite{IoT_survey}. One key challenge of the fifth generation (5G) and 5G beyond is to supporting billions of IoT devices with diversified quality of service (QoS) requirements and limited spectrum resources~\cite{IoT_survey2}. Aiming to provide connectivity for devices, two potential solutions have been proposed~\cite{IoT_solution_jsec,LpWA_mag}. On the one hand, new transmission protocol for Low-Power Wide-Area Network (LPWAN)~\cite{LPWAN_zhijin}, i.e., narrow-band Internet of Things (NB-IoT) in Release 13~\cite{3GPP_release13,NB_IoT_mag} and Long Range (LoRa) networks~\cite{LPWAN_qin}, were proposed to support connectivity requirements. On the other hand, new techniques from the existing wireless networks, i.e., non-orthogonal multiple access (NOMA), also received considerable attention for providing access services to machine-to-machine (M2M) communications or machine-type communications (MTC)~\cite{NOMA_5G_beyond_Liu,Why_NOMA}. In Long Term Evolution—-Advanced (LTE-A), a set of cellular communication protocols were proposed for MTC and IoT infrastructures~\cite{LTE-A_IoT}. In LTE-A, orthogonal multiple access (OMA) has been employed on both downlink and uplink transmission, where total transmit channel bandwidth can be partitioned into smaller bands. In the next generation IoT networks, the IoT devices located at different height attract considerable attention, e.g., aerial devices or devices located on the buildings.

In 5G and 5G beyond, unmanned aerial vehicles (UAV) or aerial devices are gaining more popularity as well as service providers or subscribers~\cite{UAV_IoT_mag}. Mozaffari~{\em et al.}~\cite{UAV_IoT_uplink} proposed a UAV assisted IoT network, where multiple UAVs play as aerial BSs for proving access services to terrestrial IoT devices. The existence of line-of-sight (LoS) link between devices and UAV platforms is probabilistic, which depends on the environment, locations of the devices and the UAVs as well as the elevation angle~\cite{3GPP_36.777}. Mei~{\em et al.}~\cite{uplink_cellular_mei} proposed a NOMA enhanced UAV communication in uplink scenario, where the small-scale fading between the UAV and devices is omitted due to the fact that the effect of path loss is the dominant component for the large scale networks. Hou~{\em et al.}~\cite{Hou_3D_UAV} proposed a NOMA enhanced UAV-to-Everything network, where UAV can provide wireless services to randomly roaming devices. The small-scale fading channels of UAV networks were discussed in~\cite{Hou_Single_UAV}, where Nakagami fading channels were employed. Hu~{\em et al.}~\cite{UAV_IoT_datacollete} proposed a UAV assisted mobile edge computing network, where UAV flies around multiple users to provide computing services. Generally speaking, Nakagami or Rician fading channels are used to evaluate the fluctuations for LoS links. It is also worth noting that the fading parameter of Nakagami fading $m=\frac{(K+1)^2}{2K+1}$, the distribution of Nakagami fading is approximately Rician fading with parameter $K$~\cite[eq. (3.38)]{wireless_communication_goldsmith}. It is estimated that by the year 2020, more than 50 billion IoT devices will be connected as components of the IoT networks~\cite{Billion_IoT}. They will generate unprecedented data, with the features of larger size, higher velocity and heterogeneity~\cite{heter_IoT}. Cloud service may be a solution for IoT networks~\cite{IoT_cloud_jsec} for significantly reducing overall power consumption. Mozaffari~{\em et al.}~\cite{Saad_UAV_cellular} proposed a 3D distributed UAV cellular network, where multiple UAVs play as aerial BSs transmitting their data to aerial users in downlink. However, given the constraint of scarce bandwidth resources, it is still challenging to serve IoT devices simultaneously in the uplink scenarios by conventional OMA techniques.

In order to solve this problem, NOMA stands as a promising solution to provide connectivity by efficiently using the available bandwidth resource~\cite{massive_NOMA_IoT,NOMA_5G_beyond_Liu}. More specifically, in contrast to the conventional OMA techniques, NOMA is capable of exploiting the available resources more efficiently by providing enhanced spectrum efficiency and connectivity on the specific channel conditions of devices~\cite{NOMA_mag_DingLiu}.
To be more clear, in NOMA enhance uplink scenarios, the BS receives the signal from multiple devices simultaneously by power domain multiplexing within the same frequency, time and code block.
The basic principles of NOMA techniques rely on the employment of successive interference cancelation (SIC) techniques at the BS~\cite{heterNOMA_Qin}, and hence multiple accessed devices can be realized in the power domain via different power levels for the BS in the same resource block. The potentials and limitations of NOMA assisted IoT networks were discussed in~\cite{Massive_IoT_NOMA_mag}, which indicates that the NOMA assisted IoT network is more efficient for the case of low target rate scenarios compared with conventional orthogonal multiple access techniques, i.e., time-division multiple access (TDMA) and frequency-division multiple access (FDMA).
Wu~{\em et al.}~\cite{IoT_NOMAorTDMA} proposed a NOMA enhanced wireless powered IoT network. The performance gap between NOMA and OMA enhanced wireless powered IoT network has been compared. Furthermore, Zhai~{\em et al.}~\cite{NOMA_EE} optimized energy-efficiency in a NOMA enhanced multi-device IoT network. Ding~{\em et al.}~\cite{MIMO-NOMA_IoT_Ding} proposed a multiple-input multiple-output (MIMO)-NOMA design for IoT transmission, where two IoT devices are grouped to perform NOMA. Moon~{\em et al.}~\cite{SCMA_random_access} proposed a sparse code multiple access enhanced IoT network, where multiple randomly roaming devices are connected to the BS. Lv~{\em et al.}~\cite{MiliWave_M2M_downlink_Lv} proposed a NOMA enhanced IoT network in millimeter-wave transmission, where the system performance has been evaluated in downlink transmission. Shao~{\em et al.}~\cite{Fog_IoT_NOMA} proposed a hybrid NOMA enhanced fog computing network. The device clustering and power allocation strategies were optimized. A NOMA assisted IoT network was proposed for the case that IoT devices have strict latency requirements and no retransmission opportunities are available~\cite{massive_NOMA_grantFree}. Shirvanimoghaddam~{\em et al.}~\cite{Jasc_massive_IoT_cellular} proposed a IoT scenario in cellular networks, where the throughput and energy efficiency in a NOMA scenario with random packets arrival model were evaluated.

Previous contributions related to NOMA networks mainly focus on the two-user to four-user cases~\cite{massive_NOMA_ISIC,massive_NOMA_relay}. In order to provide connectivity to multiple devices simultaneously, a novel clustering strategy based on stochastic geometry tools is proposed, where multiple devices can be simultaneously served by utilizing NOMA technique, and the BS can simply decode the signal of devices from the nearest device to the farthest device. In practice, obtaining the CSI at the transmitter or receiver is not a trivial problem, which requires the classic pilot-based training process. Thus, it is not possible to evaluate the accurate CSI for devices due to the unacceptable computational complexity.
To-date, to the best of our knowledge, there has been no existing research contribution intelligently investigating the performance of NOMA enhanced IoT networks, particularly with the focus of 3-D distributed devices, which motivates us to develop this treatise. NOMA enhanced terrestrial and aerial networks design has to tackle three additional challenges: i) Having NOMA devices imposes additional intra-pair interference at the BS; ii) The aerial network has to consider different fading channels to evaluate the gain of LoS/NLoS link; iii) The connected devices dramatically increase the analyse complexity.
In this article, aiming at tackling the aforementioned issues, we propose a NOMA enhanced IoT network, where only partial CSI, distance information, is required to cluster multiple devices. It is also worth noting that the proposed NOMA network is a good solution for the delay sensitive IoT devices. The transmission can be started after synchronize immediately.

\subsection{Contributions}

In contract to most existing research contributions in context of NOMA enhanced IoT networks~\cite{IoT_NOMAorTDMA,MIMO-NOMA_IoT_Ding,SCMA_random_access,MiliWave_M2M_downlink_Lv,massive_NOMA_grantFree,Jasc_massive_IoT_cellular}, where the CSI is perfectly known at the BS. We consider a novel NOMA enhanced network, where only partial CSI is required\footnote{Generally speaking, the packet length is finite for the IoT networks, which results in an additional decoding error probability at receivers~\cite{jsac_finite_blocklength,high_cited_small_packets}. However, the small packet length for the IoT networks does not significantly affect the accuracy of numerical analysis. Thus, we neglect it in this article for simplicity.}. Based on the proposed network, the primary theoretical contributions can be summarized as follows:

\begin{itemize}
  \item We develop a novel clustering strategy for the NOMA enhanced IoT networks, where only distance information is required to cluster devices. We then develop two potential scenarios to address the impact of NOMA on the network performance, where stochastic geometry approaches are invoked to model the locations of both aerial and terrestrial devices.
  \item \emph{For the NOMA enhanced terrestrial networks}: we derive the new channel statistics for terrestrial and aerial devices. The closed-form expressions of clustered devices in terms of coverage probability are derived. Additionally, we derive the general expressions in terms of coverage probability for the OMA enhanced terrestrial IoT networks. Our analytical results illustrate that the coverage probability of the far devices depends on the nearer devices.
  \item \emph{For the NOMA enhanced aerial networks}: we derive the exact analytical expressions of NOMA users in terms of coverage probability. The asymptotic coverage probabilities are derived. Our analytical results illustrate that it is more preferable to cluster far devices with NLoS links.
  \item Simulation results confirm our analysis, and illustrate that by setting coverage radius and targeted rate properly, the proposed NOMA enhanced network has superior performance over OMA enhanced network in terms of coverage probability, which demonstrates the benefits of the proposed strategies. Our analytical results also illustrate that the proposed NOMA enhanced network is not in need of a larger transmit power for increasing the coverage probability due to the fact that the coverage probability ceiling occurs in the high SNR regime. For the case of finite packet length, it is demonstrated that the impacts of packet length on the achievable rate are getting stronger with increased number of devices.
\end{itemize}

\subsection{Organization and Notations}
The rest of this article is organized as follows. In Section \uppercase\expandafter{\romannumeral2}, both the NOMA enhanced terrestrial and aerial networks are investigated, where the BS provides access services to the terrestrial or aerial devices located in the different power zones. In Section \uppercase\expandafter{\romannumeral3}, the coverage performance of the proposed network is investigated. Our numerical results are demonstrated in Section~\uppercase\expandafter{\romannumeral4} for verifying our analysis, which is followed by the conclusion in Section \uppercase\expandafter{\romannumeral5}.

\section{System Model}

\begin{figure}[t!]
\centering
\includegraphics[width =3.3in]{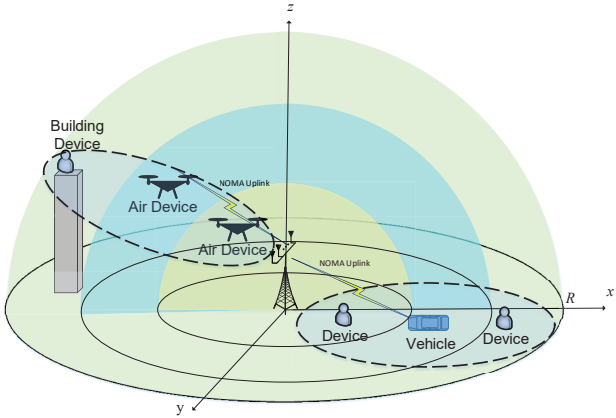}
\caption{Illustration of a typical NOMA enhanced IoT network supported by an omni-antenna.}
\label{system_model}
\end{figure}

Consider a NOMA enhanced uplink communication scenario in which multiple terrestrial and aerial devices equipped with a single omni transmitting antenna each are communicating with a BS equipped with a single omni receiving antenna. Fig.~\ref{system_model} illustrates the NOMA enhanced wireless communication model with a single BS.

The terrestrial devices are located in the different power zones according to homogeneous Poisson point process (HPPP), which is denoted by $\Psi_g$ and associated with the density $\lambda_g$. It is assumed that $M$ terrestrial devices transmit their signal to the BS via NOMA protocol, where $M$ devices are located in different power zones. Without loss of generality, the disc $\mathcal{R}^2$ with the radius $R$ is equally separated to $M$ different power zones by distance, e.g., the radius of the $i$-th power zone is between $\frac{{(i - 1)R}}{M}$ to $\frac{{iR}}{M}$ for the case of $1<i\le M$. In this article, we define the device located in the $i$-th power zone as user $i$.

Consider the use of a composite channel model with two parts, large-scale fading and small-scale fading. $L$ denotes the large-scale fading between the BS and devices. It is assumed that large-scale fading and small-scale fading are independently and identically distributed (i.i.d.). In this article, large-scale fading represents the path loss between the BS and devices, which can be expressed as
\begin{equation}\label{large scale fading_ground,eq1}
{L_{g,i}}(d) = \left\{ {\begin{array}{*{20}{l}}
{d_{g,i}^{ - {\alpha _g}},\,if\,\,{d_{g,i}} > {r_0}}\\
{r_0^{ - {\alpha _g}},\,otherwise}
\end{array}} \right.,
\end{equation}
where $d_{g,i}$ denotes the distance between the BS and terrestrial device $i$, and $\alpha_g$ denotes the path loss exponent for terrestrial devices. The parameter $r_0$ avoids a singularity when the distance is small. For simplicity, it is assumed that the minimum radius of the disc and space are greater than $r_0$.

Due to the fact that the strong scattering between the BS and terrestrial devices, the small-scale fading of device $i$ is defined by Rayleigh fading, which is denoted by ${\left| {{h_{g,i}}} \right|^2}$, and the probability density functions (PDFs) can be expressed as
\begin{equation}\label{channel matrix rayleigh,eq3}
{f}(x) = {e^{ - {{x}}}}.
\end{equation}
Since large-scale fading is the dominate component of attenuations, the BS only needs partial CSI, the distance information between devices and the BS, to group multiple devices in a NOMA cluster. Thus, given the channel gain relationship of multiple devices, we have $d_{g,1}^{ - {\alpha _g}}{\left| {{h_{g,1}}} \right|^2} > d_{g,2}^{ - {\alpha _g}}{\left| {{h_{g,2}}} \right|^2} >  \cdots  > d_{g,M}^{ - {\alpha _g}}{\left| {{h_{g,M}}} \right|^2}$ at the BS. In this article, it is assumed that the transmit power of multiple devices are the same, and therefore the BS can decode multiple devices from the nearest device to the farthest device.

We then turn our focus on the system model of aerial IoT networks. For tractability purpose, the coverage space of NOMA enhanced aerial network is a semi-sphere, denoted by $\mathcal{V}^3$, and the radius of the sphere is $R$. Without loss of generality, we also assume that the space $\mathcal{V}^3$ is equally separated to $M$ different power spaces according to the distance, where $M$ aerial devices are uniformly distributed in the different power spaces according to HPPP, which is denoted by $\Psi_a$ and associated with the density $\lambda_a$. It is assumed that in the association step, $M$ aerial devices transmit their signal to the BS via NOMA protocol, and the large-scale fading of aerial device $i$ can be expressed as
\begin{equation}\label{large scale fading_UAV,eq1}
{L_{u,i}}(d) = \left\{ {\begin{array}{*{20}{l}}
{d_{u,i}^{ - {\alpha _u}},\,if\,\,{d_{u,i}} > {r_0}}\\
{r_0^{ - {\alpha _u}},\,otherwise}
\end{array}} \right.,
\end{equation}
where $\alpha_u$ denotes the path loss exponent for aerial devices. Note that in Cartesian coordinates, in order to evaluate the distance between the BS and aerial devices, the horizontal distance, vertical distance, and altitude are separated components. On the contrary, in polar coordinates, we generally focus on the overall distance, horizontal angle and vertical angle.

In order to further illustrate the LoS links between the BS and aerial devices, the small-scale fading is defined by Nakagami fading, and the PDFs can be written as
\begin{equation}\label{channel matrix,eq3}
{f}(x) = \frac{{m^m {x^{{m} - 1}}}}{{\Gamma ({m})}}{e^{ - {{mx}}}},
\end{equation}
where $m$ denotes the fading parameter of the environment, and $\Gamma ({m})$ represents the Gamma function. Note that $\Gamma ({m})=(m-1)!$ when $m$ is an integer. It is worth noting that for the case of $m=1$, Nakagami fading channel degrades to Raleigh fading channel. Generally speaking, stronger fading environment results in higher fading parameter in Nakagami fading channels. For notation simplicity, $\left| h_{u,i} \right|^2$ denotes the small-scale channel coefficient for aerial device $i$. Thus, similar to the terrestrial networks, the decoding at the BS starts from the nearest aerial device to the farthest aerial device one by one.

In uplink transmission, the BS receives the signal for multiple terrestrial and aerial devices simultaneously. Thus, the received power for the BS is given by
\begin{equation}\label{received user power}
{P_B} = \sum\limits_{i = 1}^M {{P_g}} {L_{g,i}}{\left| {{h_{g,i}}} \right|^2} + \sum\limits_{i = 1}^M {{P_u}} {L_{u,i}}{\left| {{h_{u,i}}} \right|^2}+ {\sigma ^2},
\end{equation}
where ${\sigma ^2}$ denotes the additive white Gaussian noise (AWGN) power, $P_g$ and $P_u$ denote the transmit power of terrestrial and aerial devices, respectively. Note that the proposed design cannot guarantee the optimal performance for the NOMA enhanced network. More sophisticated designs on transmit power levels can be developed for further enhancing the attainable performance of the network considered, but this is beyond the scope of this treatise. Besides, it is assumed that the CSI of both terrestrial and aerial devices are partly known, where the information of small-scale fading is unknown at the BS.

\section{NOMA Enhanced IoT Networks}
\subsection{NOMA Enhanced Terrestrial IoT Networks}

We first discuss the performance of the NOMA enhanced terrestrial IoT networks. New channel statistics and coverage probabilities are illustrated in the following subsections.

\subsubsection{New Channel Statistics}

In this subsection, we derive new channel statistics for the NOMA enhanced networks, which will be used for evaluating the coverage probabilities in the following subsections.
\begin{lemma}\label{lemma1:new channel statistics for ground devices}
Assuming that terrestrial devices are i.i.d. located according to HPPPs in the disc $\mathbb{R}^2$ of Fig.~\ref{system_model}. In order to provide access services to devices simultaneously by NOMA technique, multiple users located in different power zones are grouped. Therefore, the PDFs of terrestrial device $i$ can be given by
\begin{equation}\label{PDF of terrestrial devices}
{f_{g,i}}(r) = \left\{ \begin{array}{l}
\frac{{2{M^2}r}}{{{R^2} - {M^2}r_0^2}},\,\,i = 1,{r_0} < r < \frac{R}{M}\\
\frac{{2{M^2}r}}{{(2i-1){R^2}}},\,1 < i \le M,\frac{{(i - 1)R}}{M} < r < \frac{{iR}}{M}
\end{array} \right.,
\end{equation}
where $M \ge 2$.
\begin{proof}
We first focus on the nearest device, who is located in the disc with the radius $r_o$ to $\frac{R}{M}$. According to HPPP, the PDF of the nearest device can be derived by
\begin{equation}\label{PDF of terrestrial devices defination, first device}
{f_{g,1}}\left( r \right) = \frac{{{\lambda _g}{\Psi _g}2\pi r}}{{{\lambda _g}{\Psi _g}\left( {\pi {{\left( {\frac{{R}}{M}} \right)}^2} - \pi {{\left( {r_o} \right)}^2}} \right)}}.
\end{equation}

Again, according to HPPPs, the PDF of terrestrial devices $i$ can be given by
\begin{equation}\label{PDF of terrestrial devices defination}
{f_{g,i}}\left( r \right) = \frac{{{\lambda _g}{\Psi _g}2\pi r}}{{{\lambda _g}{\Psi _g}\left( {\pi {{\left( {\frac{{iR}}{M}} \right)}^2} - \pi {{\left( {\frac{{(i - 1)R}}{M}} \right)}^2}} \right)}},
\end{equation}
if $i>1$. After some algebraic manipulations, the proof of Lemma~\ref{lemma1:new channel statistics for ground devices} is complete.
\end{proof}
\end{lemma}

We then turn our attention to the aerial devices.
It is assumed that the aerial devices are uniformly located in the coverage space $\mathbb{V}^3$ within the difference power spaces, and thus the PDFs of aerial devices can be given in the following Lemma.

\begin{lemma}\label{lemma2:new channel statistics for aerial devices}
Assuming that aerial devices are i.i.d. located according to HPPPs in the space $\mathbb{V}^3$ of Fig.~\ref{system_model}. The PDFs of aerial devices can be given by
\begin{equation}\label{PDF of aerial devices}
{f_{u,i}}(r) = \left\{ \begin{array}{l}
\frac{{3{M^3}{r^2}}}{{{R^3} - {M^3}r_0^2}},\,\,i = 1,{r_0} < r < \frac{R}{M}\\
\frac{{3{M^3}{r^2}}}{{{R^3}(3{i^2} - 3i + 1)}},\,1 < i \le M,\frac{{(i - 1)R}}{M} < r < \frac{{iR}}{M}
\end{array} \right. .
\end{equation}
\begin{proof}
According to HPPPs, the PDFs of the aerial devices can be given by
\begin{equation}\label{PDF of aerial devices definition,first device}
{f_{u,1}}\left( r \right) = \frac{{\frac{1}{2}}{{\lambda _u}{\Psi _u}4\pi {r^2}}}{{\frac{1}{2} {\lambda _u}{\Psi _u}\left( {\frac{4}{3}\pi {{\left( {\frac{{R}}{M}} \right)}^3} - \frac{4}{3}\pi {{\left( {r_o} \right)}^3}} \right)}},
\end{equation}
if $i=1$, and
\begin{equation}\label{PDF of aerial devices definition}
{f_{u,i}}\left( r \right) = \frac{{\frac{1}{2}}{{\lambda _u}{\Psi _u}4\pi {r^2}}}{{\frac{1}{2} {\lambda _u}{\Psi _u}\left( {\frac{4}{3}\pi {{\left( {\frac{{iR}}{M}} \right)}^3} - \frac{4}{3}\pi {{\left( {\frac{{(i - 1)R}}{M}} \right)}^3}} \right)}},
\end{equation}
if $i>1$.
After some algebraic manipulations, Lemma~\ref{lemma2:new channel statistics for aerial devices} is proved.
\end{proof}
\end{lemma}

\subsubsection{Coverage Probability}

In this subsection, we derive the coverage probability for terrestrial devices. The coverage probability is defined as the probability that the BS can successfully decode the multiplexed signal via SIC technique with a certain pre-determined SINR threshold. As such, the coverage probability for the device $i$ is given in the following Lemma.

\begin{lemma}\label{lemma3:massive NOMA coverage define}
For the proposed NOMA enhanced network with $M$ devices, the overall transmission coverage probability for device $i$ with $M\geq i$ is given by
\begin{equation}\label{i-th coverage overall}
{{\rm{P}}_{g,i,cov}}({\tau _i}) =  \prod\limits_{b = 1}^{i} {{{\rm{P}}_{g,b}}({\tau _b})} ,
\end{equation}
where ${{{\rm{P}}_{g,b}}({\tau _b})}$ denotes the coverage probability for decoding the signal of device $b$ .
\end{lemma}

\begin{remark}\label{remark1:coverage probability inherit}
The results in \eqref{i-th coverage overall} illustrate that the coverage probability of device $i$ is depending on the devices located nearer than the device $i$.
\end{remark}

\begin{remark}\label{remark2:coverage probability inherit for i, numerrical}
The results in~\eqref{i-th coverage overall} indicate that if the decoding for device $b$ with $b<i$ is failed, the coverage probability of device $i$ is zero.
\end{remark}

We then focus on analyzing the coverage probability for decoding the signal of terrestrial device $i$, which can be expressed as
\begin{equation}\label{coverage expression general}
{{\rm{P}}_{g,i}}({\tau _i}) = \int {{f_{g,i}}\left( r \right)} Pr\left\{ {Blo{g_2}\left( {1{\rm{ + }}SIN{R_{g,i}}} \right) > R_i} \right\}dr  ,
\end{equation}
where $B$ denotes the bandwidth of terrestrial device $i$, and the SINR threshold can be given by ${\tau _i}{\rm{ = }}{2^{\frac{{{R_i}}}{{\rm{B}}}}} - 1$, $R_i$ represents the target rate of the device $i$. Thus, the SINR of terrestrial device $i$ can be expressed as
\begin{equation}\label{SINR_terrestrial_i-th}
SIN{R_{g,i}} = \frac{{{P_g}d_{g,i}^{ - {{\alpha _g}} }{{\left| {{h_{g,i}}} \right|}^2}}}{{\sum\limits_{c = i + 1}^M {{P_g}d_{g,c}^{ - {{\alpha _g}} }{{\left| {{h_{g,c}}} \right|}^2}} + \sum\limits_{a = 1 }^M {{P_u}d_{u,a}^{ - {{\alpha _u}} }{{\left| {{h_{u,a}}} \right|}^2}+ {\sigma ^2}} }}.
\end{equation}

Based on \eqref{coverage expression general} and~\eqref{SINR_terrestrial_i-th}, one can obtain
\begin{equation}\label{cov probability Pr}
\begin{aligned}
{{\rm{P}}_{g,i}}({\tau _i}) = &Pr\left\{ {{{\left| {{h_{g,i}}} \right|}^2} > \frac{{{\tau _i}{\sigma ^2}}}{{{P_g}}}d_{g,i}^{\alpha _g} + \frac{{{\tau _i}d_{g,i}^{\alpha _g} }}{{{P_g}}}{I_{{\rm{g}}}} +\frac{{{\tau _i}d_{g,i}^{\alpha _g} }}{{{P_g}}}{I_{{\rm{u}}}} } \right\}\\
& {\rm{ = }}{e^{ - {\rho_i}{\sigma ^2}}}{{\mathcal{L}}_{{g,i}}}\left( {{\rho_i}} \right) {{\mathcal{L}}_{{u}}}\left( {{\rho_i}} \right) ,
\end{aligned}
\end{equation}
where ${\rho_i } = \frac{{{\tau _i}{r^{\alpha _g} }}}{{{P_g}}}$, ${I_{{\rm{g}}}}{\rm{ = }}\sum\limits_{c = i + 1}^M {{P_g}d_{g,c}^{ - {\alpha _g} }{{\left| {{h_{g,c}}} \right|}^2}} $, ${I_{{\rm{u}}}}{\rm{ = }}\sum\limits_{a = 1 }^M {{P_u}d_{u,a}^{ - {\alpha _u} }{{\left| {{h_{u,a}}} \right|}^2}} $, ${{\mathcal{L}}_{{{g,i}}}}\left( {{\rho_i }} \right)$ and ${{\mathcal{L}}_{{{u}}}}\left( {{\rho_i }} \right)$ are the Laplace transform of the power density distributions of interference from the terrestrial and aerial devices, respectively.

We then turn our attention to obtaining the Laplace transform of intra-pair interference in~\eqref{cov probability Pr}.
\begin{lemma}\label{lemma3:general expression interference for i-th IoT}
Assuming that $M$ terrestrial devices are i.i.d. located according to HPPPs in the disc $\mathbb{R}^2$. The Laplace transform of terrestrial interference for terrestrial device $i$ can be given by
\begin{equation}\label{Laplace transform general}
\begin{aligned}
&{{\mathcal{L}}_{g,i}}\left( {s} \right) = \\
& \prod\limits_{c = i + 1}^M {} \frac{1}{{(2c - 1)}}\left( {{c^2}{}_2{{\rm{F}}_1}\left( {1, - \delta_g ;1 - \delta_g ; - s{P_g}{{\left( {\frac{M}{{cR}}} \right)}^{\alpha _g} }} \right) } \right.\\
& - \left.{ {{(c - 1)}^2}{}_2{{\rm{F}}_1}\left( {1, - \delta_g ;1 - \delta_g ; - s{P_g}{{\left( {\frac{M}{{(c - 1)R}}} \right)}^{\alpha _g} }} \right)} \right),
\end{aligned}
\end{equation}
where $\delta_g=\frac{2}{{\alpha _g}}$, and ${}_2{{\rm{F}}_1}\left( { \cdot , \cdot ; \cdot ; \cdot } \right)$ represents the Gauss hypergeometric function~\cite[eq. (3.194.2)]{Table_of_integrals}.
\begin{proof}
Please refer to Appendix A.
\end{proof}
\end{lemma}

In order to provide further engineering insight, we also provide a special case, where two terrestrial devices share the same spectrum resource simultaneously in the following Corollary, i.e., $M=2$. It is important to note that considering two users is a practical assumption which is also considered by 3rd generation partnership project (3GPP)~\cite{3GPP_NOMA_two_user}.
\begin{corollary}\label{Corollary 2 users laplace}
Assuming that two terrestrial devices are i.i.d. located according to HPPPs in the disc $\mathbb{R}^2$. The Laplace transform of terrestrial interference can be obtained in closed-form expression as
\begin{equation}\label{Corollary 2 equation}
\begin{aligned}
{\mathcal{L}}\left( {s} \right) &= \frac{4}{3}{}_2{{\rm{F}}_1}\left( {1, - \delta_g ;1 - \delta_g ; - s{P_g}{R^{ - {\alpha_g} }}} \right) \\
&- \frac{1}{3}{}_2{{\rm{F}}_1}\left( {1, - \delta_g ;1 - \delta_g ; - \frac{{s{P_g}{2^{\alpha_g} }}}{{{R^{\alpha_g} }}}} \right).
\end{aligned}
\end{equation}
\begin{proof}
By substituting $M=i=2$, the result in~\eqref{Corollary 2 equation} can be readily obtained.
\end{proof}
\end{corollary}

We then focus our attention on the Laplace transform of aerial devices in the following Lemma.
\begin{lemma}\label{lemma4: aerial device general expression interference for i-th IoT}
Assuming that $M$ aerial devices are i.i.d. located according to HPPPs in the space $\mathbb{V}^3$. The Laplace transform of aerial interference for terrestrial device $i$ can be given by
\begin{equation}\label{Laplace transform general in lemma for aerial device}
\begin{aligned}
&{{\mathcal{L}}_{u}}\left( {s} \right) = \prod\limits_{a = 1}^M {} \frac{{3{M^3}{{\left( { - \frac{{s{P_u}}}{{{m_a}}}} \right)}^{{\delta _u}}}}}{{{R^3}(3{a^2} - 3a + 1)( - {\alpha _u})}} \\
& \times \left( {B\left( { - \frac{{s{P_u}}}{{{m_a}}}{{\left( {\frac{M}{{aR}}} \right)}^{{\alpha _u}}}; - {\delta _u},1 - {m_a}} \right) }\right. \\
& - \left.{ B\left( { - \frac{{s{P_u}}}{{{m_a}}}{{\left( {\frac{M}{{(a - 1)R}}} \right)}^{{\alpha _u}}}; - {\delta _u},1 - {m_a}} \right)} \right),
\end{aligned}
\end{equation}
where $\delta_u=\frac{3}{{\alpha _u}}$, $m_a$ denotes the fading parameter of aerial device $a$, and ${{\rm{B}}}\left( { \cdot  ; \cdot , \cdot } \right)$ represents the incomplete Beta function~\cite[eq. (8.391)]{Table_of_integrals}.
\begin{proof}
Please refer to Appendix B.
\end{proof}
\end{lemma}

Based on derived results in~\textbf{Lemma~\ref{lemma3:general expression interference for i-th IoT}} and~\textbf{Lemma~\ref{lemma4: aerial device general expression interference for i-th IoT}}, we can obtain the coverage probability in the following Theorem.
\begin{theorem}\label{general theorem of coverage probability}
Assuming that the devices are located in the different power zones according to HPPPs, the coverage probability of terrestrial device $i$ can be expressed as follows:
\begin{equation}\label{theorem 1 i>1}
{{\rm{P}}_{g,i}}({\tau _i}) = \frac{{2{M^2}}}{{(2i - 1){R^2}}}\int_{\frac{{(i - 1)R}}{M}}^{\frac{{iR}}{M}} {} r{e^{ - {\rho _i}{\sigma ^2}}}{{\mathcal{L}}_{g,i}}\left( {{\rho _i}} \right) {{\mathcal{L}}_{u}}\left( {{\rho _i}} \right)dr,
\end{equation}
for $i \ge 2$, and
\begin{equation}\label{theorem 1 i=1}
{{\rm{P}}_{g,1}}({\tau _1}) = \frac{{2{M^2}}}{{(R^2 - {M^2} {r_0^2}){}}}\int_{r_0}^{\frac{{R}}{M}} {} r{e^{ - {\rho _1}{\sigma ^2}}}{{\mathcal{L}}_{g,1}}\left( {{\rho _1}} \right) {{\mathcal{L}}_{u}}\left( {{\rho _1}} \right)dr,
\end{equation}
for $i=1$.
\begin{proof}
Based on the derived results in \textbf{Lemma~\ref{lemma3:general expression interference for i-th IoT}} and \textbf{Lemma~\ref{lemma4: aerial device general expression interference for i-th IoT}}, we can first express the coverage probability of terrestrial device $i$ as follows:
\begin{equation}\label{coverage befor polar}
{{\rm{P}}_{g,i}}({\tau _i}) = \int\limits_{{\mathbb{ R}_i^2}} {} {f_{g,i}}(d_{g,i}){e^{ - {\rho _{i,g}}{\sigma ^2}}}{{\cal L}_{g,i}}\left( {{\rho _{i,g}}} \right){{\cal L}_u}\left( {{\rho _{i,g}}} \right)d(d_{g,i}^{{\alpha _g}}),
\end{equation}
where ${\rho _{i,g}} = \frac{{{\tau _i}d_{g,i}^{{\alpha _g}}}}{{{P_g}}}$. For simplicity, ${\mathbb{ R}_i^2}$ represents the ring for terrestrial devices $i$.
Upon changing to polar coordinates, we can obtain the desired results in~\eqref{theorem 1 i>1} and~\eqref{theorem 1 i=1}. Thus, the proof is complete.
\end{proof}
\end{theorem}

It is hard to obtain engineering insights from~\eqref{theorem 1 i>1} and~\eqref{theorem 1 i=1} directly, and thus we derive the following corollary.
\begin{corollary}\label{general corollary of coverage probability terrestrial users}
Assuming that the devices are located in the different power zones according to HPPPs, and $r_0 << \frac{R}{M}$, the coverage probability of device $i$ can be approximated to
\begin{equation}\label{Gauss chebechev quadra in corollary for terrestrial users}
\begin{aligned}
& {{\rm{P}}_{g,i}}({\tau _i}) \\
& \approx \frac{{M}}{{(2i - 1)R}}{\omega _n}\sum\limits_{n = 1}^N {} {\xi _n}{l_n}{e^{ - {\rho _{n,g}}{\sigma ^2}}}{{\mathcal{L}}_{g,i}}\left( {{\rho _{n,g}}} \right) {{\mathcal{L}}_{u}}\left( {\rho _{n,g}} \right),
\end{aligned}
\end{equation}
for $i > 1$, where ${\omega _n} = \frac{\pi }{N}$, ${\xi _n} = \sqrt {1 - \nu _n^2} $, ${\nu _n} = \cos \left( {\frac{{2n - 1}}{{2N}}\pi } \right)$, ${l_n} = \frac{{R({\nu _n} + 2i - 1)}}{{2M}}$, ${\rho _{n,g}}= \frac{{{\tau _i}{l_n^{\alpha_g} }}}{{{P_g}}} $, $N$ denotes  the Gaussian-Chebyshev parameter.

The coverage probability of the nearest terrestrial device can be expressed as
\begin{equation}\label{Gauss chebechev quadra in corollary for terrestrial users for i=1}
\begin{aligned}
& {{\rm{P}}_{g,1}}({\tau _1}) \\
& \approx \frac{{M}}{{R+M r_0}}{\omega _n}\sum\limits_{n = 1}^N {} {\xi _n}{l_1}{e^{ - {\rho _{1,g}}{\sigma ^2}}}{{\mathcal{L}}_{g,i}}\left( {{\rho _{1,g}}} \right) {{\mathcal{L}}_{u}}\left( {\rho _{1,g}} \right),
\end{aligned}
\end{equation}
for $i = 1$, ${l_1} = \frac{{\left( {R - M{r_0}} \right)\left( {{\nu _n} + \frac{{2R}}{{R - M{r_0}}} - 1} \right)}}{{2M}}$, ${\rho _{1,g}}= \frac{{{\tau _1}{l_1^{\alpha_g} }}}{{{P_g}}} $. Here, ${{\mathcal{L}}_{g,i}}( {{ \cdot }})$ and ${{\mathcal{L}}_{u}}( {{ \cdot }})$ are given by~\eqref{Laplace transform general} and~\eqref{Laplace transform general in lemma for aerial device}, respectively.
\begin{proof}
By utilizing Gauss-Chebyshev Quadrature, we can obtain the desired results in~\eqref{Gauss chebechev quadra in corollary for terrestrial users} and~\eqref{Gauss chebechev quadra in corollary for terrestrial users for i=1}. Thus, the proof is complete.
\end{proof}
\end{corollary}

\begin{remark}\label{remark1:radius}
The results in~\eqref{Gauss chebechev quadra in corollary for terrestrial users} demonstrate that the coverage probability is a monotonic decreasing function on the cluster radius.
\end{remark}

In order to obtain further engineering insights, we also derive the following corollary in the case of low target rate scenario.
\begin{corollary}\label{special corollary of coverage probability terrestrial users low target rate}
Assuming that the devices are located in the different power zones according to HPPPs, and $r_0 << \frac{R}{M}$. It is assumed that the target rate of terrestrial device $i$ is lower than the bandwidth, i.e., ${\tau _i}<1$, the closed-form expression in terms of coverage probability of device $i$ in the case of $P_u=0$ can be approximated to
\begin{equation}\label{corolary low tareget rate expression}
\begin{aligned}
{{\rm{P}}_{g,i}}({\tau _i}) &= \frac{{{\phi _1}\phi _2^{ - n - \delta_g }}}{\alpha_g }\left( {\gamma \left( {n + \delta_g  + 1,{\phi _2}{{\left( {\frac{{iR}}{M}} \right)}^{\alpha_g} }} \right)  } \right. \\
&- \left. { \gamma \left( {n + \delta_g  + 1,{\phi _2}{{\left( {\frac{{(i - 1)R}}{M}} \right)}^{\alpha_g} }} \right)} \right),
\end{aligned}
\end{equation}
where \begin{equation*}
      \begin{aligned}
       {\phi _1} &= \frac{{2{M^2}}}{{(2i - 1){R^2}}}\prod\limits_{m = i + 1}^M {} \frac{1}{{(2m - 1)}}\sum\limits_{n = 0}^N {\frac{{{{(1)}_n}{{( - \delta_g )}_n}}}{{{{(1 - \delta_g )}_n}n!}} }\\
        &\times {{\left( { - {\tau _i}{{\left( {\frac{M}{R}} \right)}^{\alpha_g} }} \right)}^n} {\left( {{m^{ - n{\alpha_g}  + 2}} - {{(m - 1)}^{ - n\alpha_g  + 2}}} \right)},
      \end{aligned}
      \end{equation*}
${\phi _2} = \frac{{{\tau _i}{\sigma ^2}}}{{{P_g}}}$, and ${(x)_n}$ represents rising Pochhammer symbol with ${(x)_n} = \frac{{\Gamma (x + n)}}{{\Gamma (x)}}$.
\begin{proof}
Please refer to Appendix C.
\end{proof}
\end{corollary}

In order to glean further engineering insights, the coverage probability of terrestrial device $i$ in the OMA scenario, i.e., TDMA, is also derived in the following Corollary. In the OMA scenario, multiple terrestrial devices obey the same distance distributions and small-scale fading channels. The OMA benchmark adopted in this treatise is that by dividing the multiple users in equal time/frequency slots.
\begin{corollary}\label{theorem of OMA terrestrial}
In the OMA scenario, assuming that the devices are located in the different power zones according to HPPPs, the overall coverage probability of device $i$ in the case of $P_u=0$ can be approximated as follows:
\begin{equation}\label{theorem of OMA terrestrial fomula}
\begin{aligned}
{\rm{P}}_{g,i,cov}^O(\tau _i^O) &= \frac{{2{M^2}\phi _{2,O}^{ - \delta_g }}}{{(2i - 1){R^2}{\alpha_g} }}\left( {\gamma \left( {\delta_g  + 1,{\phi _{2,O}}{{\left( {\frac{{iR}}{M}} \right)}^{\alpha_g} }} \right) }\right.\\
& - \left.{ \gamma \left( {\delta_g  + 1,{\phi _{2,O}}{{\left( {\frac{{(i - 1)R}}{M}} \right)}^{\alpha_g} }} \right)} \right),
\end{aligned}
\end{equation}
where $\tau _i^O{\rm{ = }}{2^{\frac{{M{R_i}}}{{\rm{B}}}}} - 1$, and ${\phi _{2,O}} = \frac{{\tau _i^O{\sigma ^2}}}{{{P_g}}}$.
\begin{proof}
We first derive the coverage probability expression of terrestrial device $i$ in the OMA case as follows
\begin{equation}\label{OMA SINR expression}
\Pr \left\{ {\frac{B}{M}lo{g_2}\left( {1{\rm{ + }}SIN{R_{i,O}}} \right) > {R_i}} \right\},
\end{equation}
where $SIN{R_{i,O}} = \frac{{{P_g}d_{g,i}^{ - {\alpha_g} }{{\left| {{h_{g,i}}} \right|}^2}}}{{{\sigma ^2}}}$. Following the similar steps in Appendix C, the result in~\eqref{theorem of OMA terrestrial fomula} can be readily proved.
\end{proof}
\end{corollary}

\begin{remark}\label{remark4:coverage probability of OMA}
The results in~\eqref{theorem of OMA terrestrial fomula} indicate that the coverage probability of multiple devices in the OMA scenario is independent on other devices, whereas the coverage probability of devices depend on the nearer devices in the proposed NOMA enhanced networks.
\end{remark}

We also want to provide some benchmark schemes in TABLE~\ref{required RBs}. We use “RBs” to represent the required number of resource blocks for the case that the amount number of devices is set to 1000. For the proposed NOMA networks, the required number of RBs is 200 by setting $M=5$, which indicates that the proposed NOMA network is more efficient on the RBs.

\begin{table}[h]
\centering \caption{\\ REQUIRED NUMBER OF RBs (1000 DEVICES)}
\begin{tabular}{|c|c|}
\hline
Access Mode & RBs \\
\hline
{Conventional OMA} & $1000$  \\
\hline
{SCMA~\cite{SCMA_survey}} & $667$  \\
\hline
{Proposed NOMA} & $\frac{1000}{M}$  \\
\hline
\end{tabular}
\label{required RBs}
\end{table}

\subsection{NOMA Enhanced Aerial IoT Networks}
In conventional IoT networks, the devices are located on the ground, whereas the proposed aerial networks mainly focus on providing access services to the devices with different heights, i.e., devices in buildings, UAVs regarded as terminal devices, or information collectors on the wall. The main difference of the aerial network is that the vertical angles between the BS and aerial devices, which can be transformed into the altitude, provide stronger power level of small-scale fading channels by LoS links. Based on the insights from~\cite{Nakagami_Hou}, another scenario considered in this article is the NOMA enhanced aerial IoT networks, where paired NOMA devices are located inside the coverage space as shown in Fig.~\ref{system_model}.

We first derive the SINR expression for decoding the aerial device $i$ at the BS as follows
\begin{equation}\label{SINR_aerial_i-th}
\begin{aligned}
& SIN{R_{u,i}}\\
& = \frac{{{P_u}d_{u,i}^{ - {{\alpha _u}} }{{\left| {{h_{u,i}}} \right|}^2}}}{{\sum\limits_{c = i + 1}^M {{P_u}d_{u,c}^{ - {{\alpha _u}} }{{\left| {{h_{u,c}}} \right|}^2}} + \sum\limits_{a = 1 }^M {{P_g}d_{g,a}^{ - {{\alpha _g}} }{{\left| {{h_{g,a}}} \right|}^2}+ {\sigma ^2}} }}.
\end{aligned}
\end{equation}

Thus, the coverage probability for decoding aerial device $i$ at the BS can be defined as
\begin{equation}\label{coverage expression general aerial}
{{\rm{P}}_{u,i}}({\tau _i}) = \int {{f_{u,i}}\left( r \right)} Pr\left\{ {Blo{g_2}\left( {1{\rm{ + }}SIN{R_{u,i}}} \right) > R_i} \right\}dr  .
\end{equation}
Thus, the overall coverage probability can be written as
\begin{equation}\label{i-th coverage overall, aerial}
{{\rm{P}}_{u,i,cov}}({\tau _i}) =  \prod\limits_{b = 1}^{i} {{{\rm{P}}_{u,b}}({\tau _b})}.
\end{equation}

Then we pay our attention on the coverage behavior of aerial devices. The coverage probability of the aerial devices is more complicated due to the LoS links, and hence we first derive the Laplace transform of the aerial interferences in following Lemma.
\begin{lemma}\label{lemma5: aerial device general expression interference for i-th IoT}
Assuming that $M$ aerial devices are i.i.d. located according to HPPPs in the different power spaces $\mathbb{V}^3$. The Laplace transform of aerial interference for decoding the signal of aerial device $i$ can be given by
\begin{equation}\label{Laplace transform general of aerial device in lemma for aerial device}
\begin{aligned}
&{{\mathcal{L}}_{u,i}}\left( {s} \right) = \prod\limits_{c = i+1}^M {} \frac{{3{M^3}{{\left( { - \frac{{s{P_u}}}{{{m_a}}}} \right)}^{{\delta _u}}}}}{{{R^3}(3{c^2} - 3c + 1)( - {\alpha _u})}} \\
& \times \left( {B\left( { - \frac{{s{P_u}}}{{{m_a}}}{{\left( {\frac{M}{{cR}}} \right)}^{{\alpha _u}}}; - {\delta _u},1 - {m_a}} \right) }\right. \\
& - \left.{ B\left( { - \frac{{s{P_u}}}{{{m_a}}}{{\left( {\frac{M}{{(c - 1)R}}} \right)}^{{\alpha _u}}}; - {\delta _u},1 - {m_a}} \right)} \right).
\end{aligned}
\end{equation}
\begin{proof}
Similar to Appendix B, Lemma~\ref{lemma5: aerial device general expression interference for i-th IoT} can be readily proved.
\end{proof}
\end{lemma}

Again, the terrestrial devices are interferences for aerial devices, and the Laplace transform of terrestrial devices can be given in the following Lemma.
\begin{lemma}\label{lemma6:general expression interference for i-th IoT}
Assuming that $M$ terrestrial devices are i.i.d. located according to HPPPs in the different power zones of Fig.~\ref{system_model}. The closed-form expression of Laplace transform of terrestrial devices for aerial device $i$ can be given by
\begin{equation}\label{Laplace transform general of terresrrial device for i-th aerial device}
\begin{aligned}
&{{\mathcal{L}}_{g}}\left( {s} \right) = \\
& \prod\limits_{a = 1}^M {} \frac{1}{{(2a - 1)}}\left( {{a^2}{}_2{{\rm{F}}_1}\left( {1, - \delta_g ;1 - \delta_g ; - s{P_g}{{\left( {\frac{M}{{cR}}} \right)}^{\alpha _g} }} \right) } \right.\\
& - \left.{ {{(a - 1)}^2}{}_2{{\rm{F}}_1}\left( {1, - \delta_g ;1 - \delta_g ; - s{P_g}{{\left( {\frac{M}{{(a - 1)R}}} \right)}^{\alpha _g} }} \right)} \right).
\end{aligned}
\end{equation}
\begin{proof}
Similar to Appendix A, the proof can be easily completed.
\end{proof}
\end{lemma}

Based on the Laplace transform of {\textbf {Lemma~\ref{lemma5: aerial device general expression interference for i-th IoT}}} and {\textbf {Lemma~\ref{lemma6:general expression interference for i-th IoT}}, the coverage probability in the case of interference limited case of the first aerial device can be derived in the following Theorem.

\begin{theorem}\label{theorem aerial coverage probability 2users gamma}
Assuming that two aerial devices are located in the different power spaces according to HPPPs, and the small-scale fading of the nearest aerial device follows Gamma distribution, the coverage probability for the nearest device can be approximated to
\begin{equation}\label{theorem aerial 2 users Gamma}
\begin{aligned}
&{{\rm{P}}_{u,1}}({\tau _1}) \\
&= \int\limits_{{r_o}}^{\frac{R}{2}} {} {f_{u,1}}(r){\sum\limits_{k = 0}^{m - 1} {\frac{{{{( - 1)}^k}}}{{k!}}\left[ {\frac{{{\partial ^k}\left( {{\left( {{\mathcal{L}_{u,i}}(s) + {\mathcal{L}_g}(s)} \right)}^m} \right)}}{{\partial {s^k}}}} \right]} _{s = {\rho _i}{r^\alpha }}}dr.
\end{aligned}
\end{equation}
where
\begin{proof}
Please refer to Appendix D.
\end{proof}
\end{theorem}

In order to glean further engineering insight, and based on the results in~\eqref{theorem aerial 2 users Gamma}, we can obtain the coverage probability in closed-form of the nearest device in the following Corollary in the case of $P_g=0$.
\begin{corollary}\label{corollary aerial coverage probability 2users gamma for m=2 and 3}
Assuming that two aerial devices are located in the different power spaces according to HPPPs, and the small-scale fading of the nearest aerial device follows Gamma distribution, the closed-form expression of coverage probability in the case of $m=2$ and $m=3$ can be approximated to
\begin{equation}\label{corollary aerial 2 users m=2}
\begin{aligned}
&{{\rm{P}}_{u,1}}({\tau _1}\left| {m = 2} \right.)\\
& \approx 6{\omega _n}\sum\limits_{n = 1}^N {} {\xi _n}t_n^2\left( { - {\mathcal{L}_{u,1}}{{\left( {{\rho _{1,u}}} \right)}^2} + 2{\mathcal{L}_{u,1}}\left( {{\rho _{1,u}}} \right)} \right),
\end{aligned}
\end{equation}
and
\begin{equation}\label{corollary aerial 2 users m=3}
\begin{aligned}
&{{\rm{P}}_{u,1}}({\tau _1}\left| {m = 3} \right.) \\
&\approx 6{\omega _n}\sum\limits_{n = 1}^N {} {\xi _n}t_n^2\left( {{\mathcal{L}_{u,1}}{{\left( {{\rho _{1,u}}} \right)}^3} - 3{\mathcal{L}_{u,1}}{{\left( {{\rho _{1,u}}} \right)}^2} + 3{\mathcal{L}_{u,1}}\left( {{\rho _{1,u}}} \right)} \right),
\end{aligned}
\end{equation}
where ${\omega _n} = \frac{\pi }{N}$, ${\xi _n} = \sqrt {1 - \nu _n^2} $, ${\nu _n} = \cos \left( {\frac{{2n - 1}}{{2N}}\pi } \right)$, ${t_n} = \frac{{R({\nu _n} +  1)}}{{4}}$, ${\rho _{1,u}}= \frac{{{\tau _1}{t_n^{\alpha_u} }}}{{{P_u}}} $, and ${{\mathcal{L}}_{u}}( {{\cdot}})$ is given by~\eqref{Laplace transform general of aerial device in lemma for aerial device}.
\begin{proof}
By utilizing Gauss-Chebyshev Quadrature, we can obtain the desired results in~\eqref{corollary aerial 2 users m=2} and~\eqref{corollary aerial 2 users m=3}. Thus, the proof is complete.
\end{proof}
\end{corollary}

\begin{remark}\label{remark3:LoS unneccesary}
Based on insight from \textbf{Remark~\ref{remark1:coverage probability inherit}} and derived results in \textbf{Corollary~\ref{corollary aerial coverage probability 2users gamma for m=2 and 3}}, one can known that the LoS links between the BS and device $i$ with $i>1$ decrease the system coverage probability.
\end{remark}

\subsection{NOMA Enhanced IoT Networks with Finite Packet Length}
One of the negligible advantages of the proposed NOMA enhanced IoT networks is low latency, and thus information may also conveyed in short-packets with finite blocklength. However, the finite blocklength results in a non-negligible decreasing of achievable rate at the BS~\cite{high_cited_small_packets,jsac_finite_blocklength}.
We then analyse achievable rate for the case of finite packet length.
In this article, we denote $R_{i,f}$ as the maximum achievable rate of device $i$ in the case of finite blocklength, which can be expressed as:
\begin{equation}\label{Maximum achievable rate}
{R_{i,f}} = {\log _2}\left( {1 + SIN{R_i}} \right) - \sqrt {\frac{{{V_i}}}{N_f}} \frac{{{Q^{ - 1}}({P_i})}}{{\ln 2}},
\end{equation}
where $N_f$ denotes the packet length, ${Q^{ - 1}}(x)$ represents the inverse of ${Q}(x) = \int_x^\infty  {\frac{1}{{\sqrt {2\pi } }}} \exp \left( {\frac{{ - {t^2}}}{2}} \right)dt$, $V_i$ denotes the channel dispersion with ${V_i} = 1 - {(1 + SIN{R_i})^{ - 2}}$, and $P_i$ is the overall outage probability of device $i$~\cite{small_packets_in_MIMO}.
\begin{remark}\label{remark5:achievable rate}
Based on the expression in~\eqref{Maximum achievable rate}, one can know that the packet length has impact on the achievable rate dramatically, where higher packet length results in larger achievable rate. We can also observe that for the case of $N_f \approx \infty$, the achievable rate can be maximized as ${R_{i,f}} = {\log _2}\left( {1 + SIN{R_i}} \right)$.
\end{remark}

\section{Numerical Studies}

In this section, numerical results are provided to facilitate the performance evaluation of NOMA enhanced IoT networks. Monte Carlo simulations are conducted for verifying analytical results. It is assumed that the bandwidth is $B=125$ kHz as one of the most common setting-ups for IoT networks. The power of AWGN is set to $\sigma^2=-174 + 10{\rm{log}_{10}}(B)$ dBm. It is also worth noting that LoS and NLoS scenarios are indicated by the Nakagami fading parameter $m$, where $m = 1$ for NLoS scenarios (Rayleigh fading) and $m >1$ for LoS scenarios. Without loss of generality, we use $m=2, 3$ to represent LoS scenario in Section IV. The minimum distance is $r_0=1m$.
The path loss exponents for the terrestrial and aerial devices are set to $\alpha_g=4$ and $\alpha_u=3$, respectively.
The radii of the disc and space are set to $R=1000m$.

\begin{figure}[t!]
\centering
\includegraphics[width =3.3in]{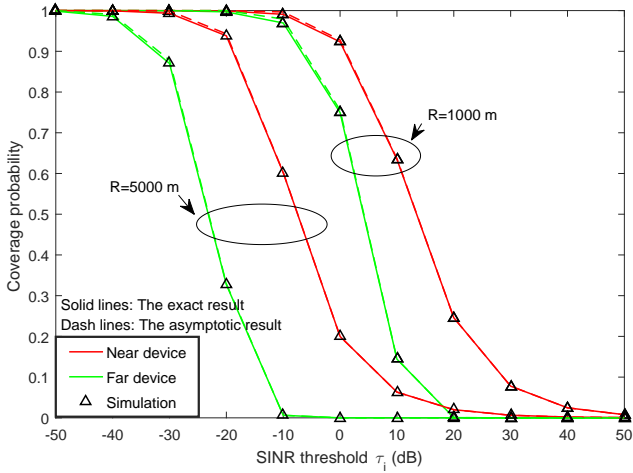}
\caption{Coverage probability of terrestrial NOMA devices versus the SINR threshold $\tau_i$ with different radius. Transmit power of terrestrial devices and aerial devices are set to $P_g=0$ dBm and $P_u=0$. The analytical and asymptotic results are derived from~\eqref{theorem 1 i>1},~\eqref{theorem 1 i=1} and~\eqref{Gauss chebechev quadra in corollary for terrestrial users}, respectively.}
\label{coverage_SINR_diff_radius}
\end{figure}

\emph{1) Impact of the Threshold and Radius:} Fig.~\ref{coverage_SINR_diff_radius} plots the coverage probability of the considered terrestrial networks with different SINR thresholds. The solid curves and dashed curves are the exact results and asymptotic results, respectively. We can see that, as the SINR threshold of the terrestrial devices increases, the coverage probability of both near and far NOMA devices decreases. This is due to fact that, as higher threshold of devices is deployed, the target rate of devices improves dramatically. It is also confirmed the closed agreement between the simulation and analytical results, which verifies our analytical results. We can also see that, the coverage probability of paired NOMA devices decreases with larger disc radius. This is due to the fact that larger disc radius increases the distance of the desired link, which also verified~\textbf{Remark~\ref{remark1:radius}}.

\begin{figure}[t!]
\centering
\includegraphics[width =3.3in]{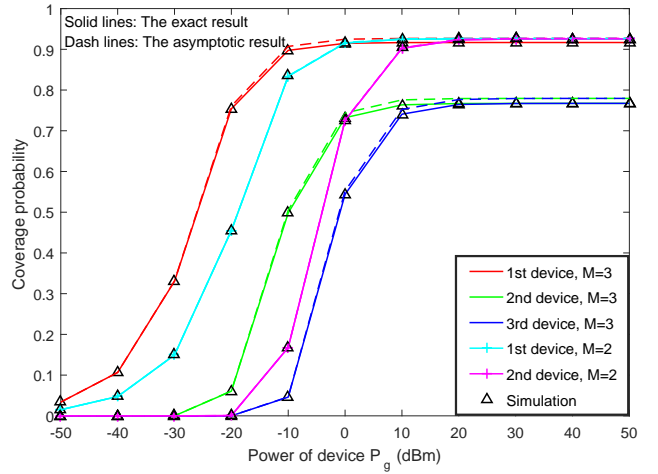}
\caption{Coverage probability of terrestrial NOMA devices versus the transmit power with different number of devices. The threshold is $\tau_i=0.5$. The analytical and asymptotic results are derived from~\eqref{theorem 1 i>1},~\eqref{theorem 1 i=1}, and~\eqref{corolary low tareget rate expression}, respectively.}
\label{coverage_3users_low_threshold}
\end{figure}

\emph{2) Impact of the Number of Devices:} Fig.~\ref{coverage_3users_low_threshold} plots the coverage probability of the considered terrestrial network with different number of accessed devices. The coverage probability of two-device case is plotted as the benchmark schemes. As we can see in the figure, coverage probability ceilings occurs, which meet the expectation due to the strong intra-pair interference. Therefore, the proposed network is not in need of a larger transmit power for increasing the coverage probability.
We can also see that the coverage probabilities for the first device and the second device in the case of $M=2$ are the same in the high SNR regime. This is due to the fact that in the high SNR regime, the coverage probability of far devices approaches one.

\begin{figure}[t!]
\centering
\includegraphics[width =3.3in]{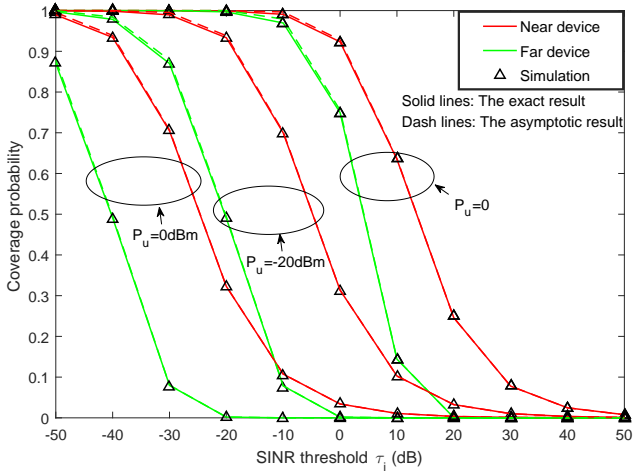}
\caption{Coverage probability of the terrestrial NOMA devices versus the SINR threshold $\tau_i$ with different power level of aerial interferences. The transmit power of terrestrial devices is $P_g=0$ dBm. The fading parameters of aerial devices are one. The analytical and asymptotic results are derived from~\eqref{theorem 1 i>1},~\eqref{theorem 1 i=1} and~\eqref{Gauss chebechev quadra in corollary for terrestrial users}, respectively.}
\label{coverage_SINR_diff_power of aerial devices}
\end{figure}

\emph{3) Impact of the Aerial Devices:} Fig.~\ref{coverage_SINR_diff_power of aerial devices} plots the coverage probability of the considered terrestrial network with different power levels of aerial devices. We can see that, the coverage probability of paired terrestrial NOMA devices decreases in the case of larger power level of aerial devices. This is due to the fact that the received interference power from aerial devices increases dramatically, which leads to the decrease of received SINR at the BS for all the terrestrial devices.

\begin{figure}[t!]
\centering
\includegraphics[width =3.3in]{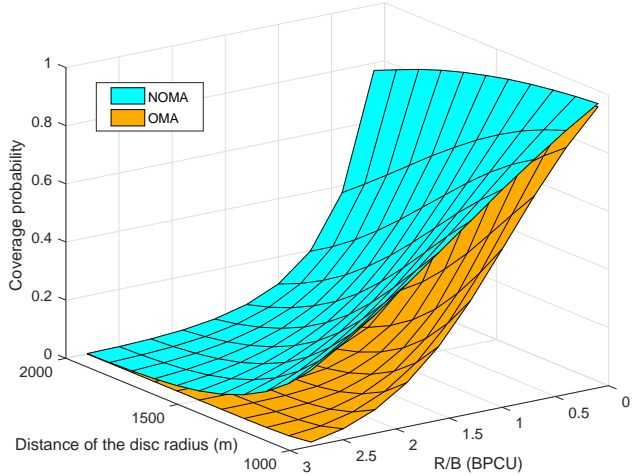}
\caption{Coverage probability of both the NOMA and OMA enhanced network versus disc radius and target rate in the case of $M=2$. The transmit power of devices is $P_g=0$ dBm.}
\label{coverage_3D_terrestrial}
\end{figure}

\emph{4) Performance Comparing with OMA:} In Fig.~\ref{coverage_3D_terrestrial}, we evaluate the coverage probability of both NOMA and OMA enhanced networks with different disc radius and target rate in two-device scenario. The coverage probability is derived by ${\rm{{P}}}_{g,1,cov} \times {\rm{{ P}}}_{g,2,cov} $. The two-device scenario of OMA enhanced network in terms of coverage probability is derived by ${\rm{P}}_{g,1,cov}^O \times {\rm{P}}_{g,2,cov}^O$. As can be seen from Fig.~\ref{coverage_3D_terrestrial}, the coverage probability of NOMA enhanced network is higher than the OMA enhanced networks, which implies that NOMA enhanced networks is capable of providing better network performance than OMA.

\begin{figure}[t!]
\centering
\includegraphics[width =3.3in]{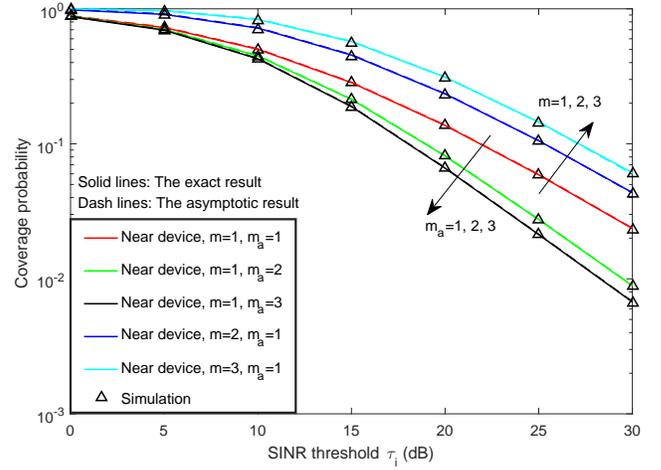}
\caption{Coverage probability of the near aerial NOMA device versus the SINR threshold $\tau_i$ with different fading parameters. The transmit power of aerial devices is $P_u=0$ dBm.}
\label{coverage_aerial_diff_fading_parameter}
\end{figure}

\emph{5) Impact of the Threshold and Fading Environments:} Fig.~\ref{coverage_aerial_diff_fading_parameter} plots the coverage probability of the considered aerial networks with different fading parameters. The fading parameters of two NOMA devices are set to $m=1, 2, 3$. On the one hand, we can see that higher fading parameter $m_a$ between the BS and farer aerial device would result in reduced coverage probability. On the other hand, higher fading parameter $m$ between the BS and the nearest device increases the coverage probability. This is because that the LoS link between the BS and aerial devices provides higher received power level.

\begin{figure}[t!]
\centering
\includegraphics[width =3.3in]{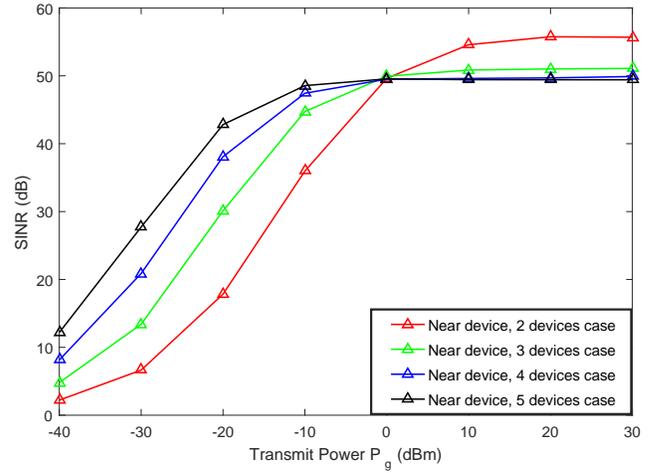}
\caption{SINR of the nearest NOMA device in the terrestrial network versus transmit power $P_g$ with different number of serving devices. The power of aerial devices is set to $P_u=0$.}
\label{SINR versus devices}
\end{figure}

\emph{6) Impact of the Number of Devices:} Fig.~\ref{SINR versus devices} plots the SINR threshold with different number of devices. On the one hand, in the low transmit power regime, the SINR performance for the nearest devices in the case of five-device case is better than the two-device case. This is because that the distance of the nearest devices in five-device case is much smaller than the two-device case. Observe that in the high transmit power regime, the SINR of two-device case is higher than other cases, which indicates that the interference of two-device case is the minimum case. Additionally, there is a cross point of curves, which indicates that there exists an optimal point for the proposed scenario.
It is also not in need of a larger transmit power for increasing the coverage probability due to the fact that the coverage ceilings occur in the high transmit power regime.

\begin{figure}[t!]
\centering
\includegraphics[width =3.3in]{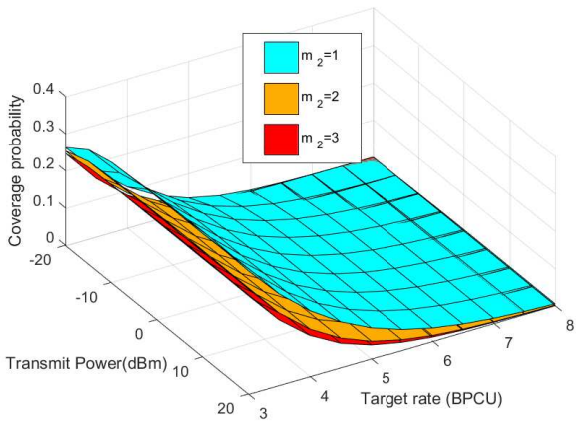}
\caption{Coverage probability of NOMA enhanced aerial network versus the transmit power and target rate in the case of $m_1=1, m_2=1, 2, 3$.}
\label{aerial_3D}
\end{figure}

\emph{7) Impact of LoS Links between the BS and Far Devices:} In Fig.~\ref{aerial_3D}, we evaluate the coverage probability of NOMA enhanced aerial networks with different fading parameter and target rate. The fading parameter of the nearest device is 1, whereas the fading parameter of the far device is $m_2=1, 2, 3$. The coverage probability is derived by ${\rm{{P}}}_{u,1,cov} \times {\rm{{ P}}}_{u,2,cov} $. As can be seen from Fig.~\ref{aerial_3D}, the coverage probability decreases when increasing fading parameters of the far device, which implies that the proposed network prefer to provide access services to the far devices with Rayleigh fading channels, which can improve the system coverage performance. This observation also verifies our \textbf{Remark~\ref{remark3:LoS unneccesary}}.

\begin{figure}[t!]
\centering
\includegraphics[width =3.3in]{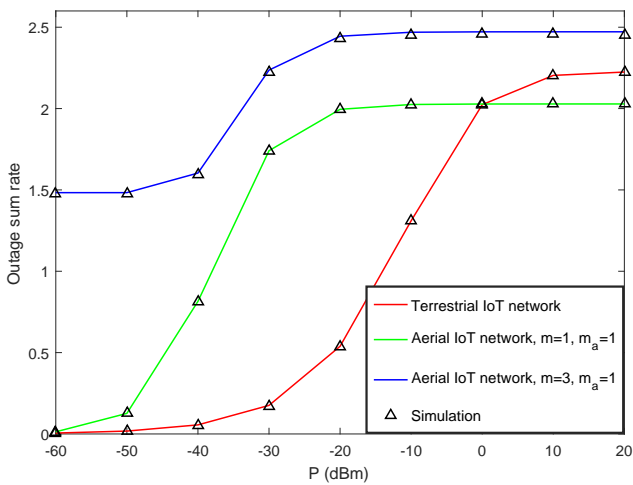}
\caption{Outage sum rate of both NOMA enhanced terrestrial and aerial network versus the transmit power with $P_u$=$P_g$=$P$ and $M=2$. The target rate of near devices and far devices are set to $R_1=1.5$ BPCU and $R_2=1$ BPCU, respectively. }
\label{Outage_sum_rate}
\end{figure}

\emph{8) Outage Sum Rate:} Fig.~\ref{Outage_sum_rate} plots the system outage sum rate versus the transmit power with different fading parameters for both terrestrial and aerial networks. The outage sum rate of terrestrial devices is derived by $R=P_{g,1,cov} \times R_1 + P_{g,2,cov} \times R_2$, where $P_{g,1,cov}$ and $P_{g,2,cov}$ denote the overall coverage probability of near devices and far devices, respectively. The outage sum rate of aerial devices is derived by $R=P_{u,1,cov} \times R_1 + P_{u,2,cov} \times R_2$.
One can observe that the case $m=3$ achieves the highest throughput since it has the lowest outage probability among the three selection fading parameters. The figure also demonstrates the existence of the throughput ceilings in the high SNR region. This is due to the fact that the coverage probability is approaching constant and the throughput is determined only by the targeted data rate. We can also see that the throughput ceilings for aerial networks in the case of $m=1, m_2=1$ is smaller than the terrestrial networks. This is due to the fact that the average distance of devices in the aerial networks is greater than the terrestrial networks, which actually decreases the received SINR for the nearest devices.

\begin{figure}[t!]
\centering
\includegraphics[width =3.3in]{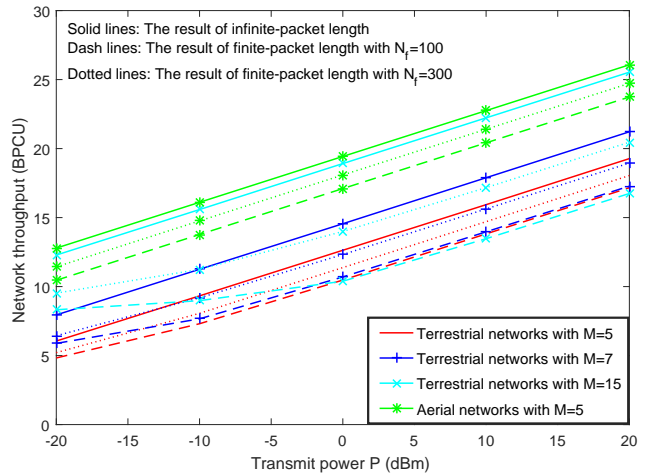}
\caption{Network throughput of both NOMA enhanced terrestrial and aerial networks versus the transmit power with $P_u$=$P_g$=$P$. The target rate of the nearest device and far devices are set to $R_1=1.5$ BPCU and $R_2, \cdots, R_M=1$ BPCU, respectively. The fading parameters of the aerial devices are set to $m_1=3$ and $m_2, \cdots, m_M =1$.}
\label{Achievable rate finite packet length}
\end{figure}

\emph{9) Achievable Throughput with Finite Packet Length:} Fig.~\ref{Achievable rate finite packet length} plots the network throughput versus the transmit power with different packet length for both terrestrial and aerial networks. The achievable rate of each device is derived by~\eqref{Maximum achievable rate}, where the system throughput is derived by the summation of multiple devices. The solid curve, dashed curve and dotted curve are the achievable rate of the infinite packet length $N_f \approx \infty$, finite packet length with $N_f=100$ and $N_f=300$, respectively. It is observed that the larger packet length is capable of providing higher achievable rate. Based on blue curves and red curves, one can observe that the gap of network throughput between infinite packet length scenario and finite packet length scenario increases, which indicates that the network throughput is sensitive on the packet length.
We can also see that the system throughput of the aerial networks is larger than the terrestrial networks. This is due to the fact that the path loss exponent in the aerial networks is smaller than the terrestrial networks, which significantly increases the network throughput. The cyan curves are the achievable network throughput for the case of $M=15$. Observe that the network throughput supported by NOMA technique are nearly the same for the case of $N_f=100$, which indicates that the network throughput cannot be enhanced for the short packet scenario. This observation shown that NOMA technique may not be a good solution for the short packet scenario. On the other hand, one can also observe that for the case of $P=10$dBm, the network throughput and average throughput in each resource block are 15.9, 17.8, 22.2 BPCU and 3.18, 2.54, 1.48 BPCU for the case of $M=5,7,15$, respectively, which indicate that the network throughput increases with the number of devices, whereas the average throughput of each user decreases. Since the data size of each device in the IoT networks is smaller than the public communications, the proposed network is more suitable for the IoT networks.

\section{Conclusions}

In this article, the application of the NOMA enhanced IoT networks was proposed. Specifically, a novel clustering strategy was proposed, where only partial CSI is required. Stochastic geometry tools were invoked for modeling the spatial randomness of both terrestrial and aerial devices. Additionally, new closed-form expressions in terms of coverage probability were derived for characterizing the network performance. The performance of OMA enhanced networks was also derived as the benchmark schemes. It was analytically demonstrated that the NOMA enhanced networks are capable of outperforming OMA enhanced networks. Based on the insights from~\cite{Saad_UAV_cellular}, one promising future direction is to accommodate UAV-BS to the massive NOMA enhanced IoT network, which is capable of proving better network performance. Furthermore, some specific scenarios for NOMA enhanced IoT networks are worth investigated, i.e., energy limited scenarios and small package scenarios.

\numberwithin{equation}{section}
\section*{Appendix~A: Proof of Lemma~\ref{lemma3:general expression interference for i-th IoT}} \label{Appendix:As}
\renewcommand{\theequation}{A.\arabic{equation}}
\setcounter{equation}{0}

Recall that the intra-pair interference received at the BS for decoding terrestrial device $i$ can be expressed as
\begin{equation}\label{appendix A interference SINR}
{I_{g,i}}{\rm{ = }}\sum\limits_{c = i + 1}^M {{P_g}d_{g,c}^{ - {\alpha_g} }{{\left| {{h_{g,c}}} \right|}^2}} .
\end{equation}
Therefore, the expectation for the intra-pair interference can be calculated as follows:
\begin{equation}\label{appendix A interference equation}
\begin{aligned}
{{\mathcal{L}}_{g,i}}\left( {s} \right) &= {\mathbb{E}}\left\{ {\exp \left( { - s\sum\limits_{c = i + 1}^M {{P_g}d_{g,c}^{ - {\alpha_g} }}{{\left| {{h_{g,c}}} \right|}^2} } \right)} \right\} \\
&  = {\mathbb{E}} \left\{ {\prod\limits_{c = i + 1}^M {} \exp \left( { - s{P_g}d_{g,c}^{ - {\alpha_g} }{{\left| {{h_{g,c}}} \right|}^2} } \right)} \right\}  \\
&  \mathop  =   \limits^{(a)} \int\limits_0^\infty  {\prod\limits_{c = i + 1}^M {} \exp \left( { - s{P_g}d_{g,c}^{ - {\alpha_g} }} \right)\exp ( - x)} dx  \\
&  = {\mathbb{E}} \left\{ {\prod\limits_{c = i + 1}^M {\frac{1}{{1 + s{P_g}d_{g,c}^{ - {\alpha_g} }}}} } \right\}.
\end{aligned}
\end{equation}
where (a) can be gleaned by the fact that ${\left| {{h_{g,m}}} \right|}$ follows Rayleigh distribution.

Recall that the distance PDFs of terrestrial interferences follow~\eqref{PDF of terrestrial devices}, and thus the Laplace transform can be transformed into
\begin{equation}\label{appendix A interference integral}
\begin{aligned}
&{{\mathcal{L}}_{g,i}}\left( {s} \right)= \sum\limits_{c = i + 1}^M {} {f_{g,c}}\left( x \right)\int_{\frac{{(c - 1)R}}{M}}^{\frac{{cR}}{M}} {\frac{x}{{1 + s{P_g}x_{}^{ - {\alpha_g} }}}} dx \\
&\mathop  = \limits^{(b)}  \sum\limits_{c = i + 1}^M {\frac{{2{M^2}\left( {s{P_g}} \right)\delta_g }}{{(2c - 1){R^2}{\alpha_g} }}} \int_{s{P_g}{{\left( {\frac{{cR}}{M}} \right)}^{ - {\alpha_g} }}}^{s{P_g}{{\left( {\frac{{(c - 1)R}}{M}} \right)}^{ - {\alpha_g} }}} {\frac{{{t^{ - \delta_g  - 1}}}}{{1 + t}}} dt,
\end{aligned}
\end{equation}
where (b) is obtained by using $t = s{P_g}x_{}^{ - {\alpha_g} }$, and by applying~\cite[eq. (3.194.2)]{Table_of_integrals}, we can obtain the Laplace transform in an elegant form in~\eqref{Laplace transform general}. The proof is complete.

\numberwithin{equation}{section}
\section*{Appendix~B: Proof of Lemma~\ref{lemma4: aerial device general expression interference for i-th IoT}} \label{Appendix:Bs}
\renewcommand{\theequation}{B.\arabic{equation}}
\setcounter{equation}{0}

Recall that multiple aerial devices are located in the space $\mathbb{V}^3$ according to HPPPs, and thus the Laplace transform of aerial interference can be expressed as
\begin{equation}\label{lapace transform in appendix C general expression}
\begin{aligned}
{{\mathcal{L}}_{u}}\left( {s} \right) &= \mathbb{E} \left\{ {\exp \left( { - s\sum\limits_{a = 1}^M {{P_u}d_{u,a}^{ - {\alpha _u}}{{\left| {{h_{u,a}}} \right|}^2}} } \right)} \right\} \\
&  = \mathbb{E}\left\{ {\prod\limits_{a = 1}^M {} \exp \left( { - s{P_u}d_{u,a}^{ - {\alpha _u}}{{\left| {{h_{u,a}}} \right|}^2}} \right)} \right\} \\
& \mathop  = \limits^{(a)} \mathbb{E}\left\{ {\prod\limits_{a = 1}^M {{{\left( {1 + \frac{{s{P_u}d_{u,a}^{ - {\alpha _u}}}}{{{m_a}}}} \right)}^{ - {m_a}}}} } \right\},
\end{aligned}
\end{equation}
where (a) can be obtained by the fact that ${\left| {{h_{u,a}}} \right|}$ follows Nakagami distribution.

By applying the distance distribution of the aerial devices in~\eqref{PDF of aerial devices}, the above expectation can be further transformed into
\begin{equation}\label{appendix C laplace transform final}
\begin{aligned}
{{\mathcal{L}}_{u}}\left( {s} \right) &=\prod\limits_{a = 1}^M {\int_{\frac{{(a - 1)R}}{M}}^{\frac{{aR}}{M}} {} {f_{u,a}}\left( x \right){{\left( {1 + \frac{{s{P_u}x_{}^{ - {\alpha _u}}}}{{{m_a}}}} \right)}^{ - {m_a}}}} dx \\
& \mathop  = \limits^{(b)} \prod\limits_{a = 1}^M {} \frac{{3{M^3}{{\left( { - \frac{{s{P_u}}}{{{m_a}}}} \right)}^{{\delta _u}}}}}{{{R^3}(3{a^2} - 3a + 1)( - {\alpha _u})}} \\
& \times \int_{ - \frac{{s{P_u}}}{{{m_a}}}{{\left( {\frac{M}{{(a - 1)R}}} \right)}^{{\alpha _u}}}}^{ - \frac{{s{P_u}}}{{{m_a}}}{{\left( {\frac{M}{{aR}}} \right)}^{{\alpha _u}}}} {\frac{{{t^{ - {\delta _u} - 1}}}}{{{{(1 - t)}^{{m_a}}}}}} dt,
\end{aligned}
\end{equation}
where (b) is obtained by applying $t =  - \frac{{s{P_g}x_{}^{ - {\alpha _u}}}}{{{m_a}}}$, and by applying~\cite[eq. (8.391)]{Table_of_integrals}, we can obtain the Laplace transform in~\eqref{Laplace transform general in lemma for aerial device}. The proof is complete.

\numberwithin{equation}{section}
\section*{Appendix~C: Proof of Corollary~\ref{special corollary of coverage probability terrestrial users low target rate}} \label{Appendix:Cs}
\renewcommand{\theequation}{C.\arabic{equation}}
\setcounter{equation}{0}

We first expand Gauss hypergeometric function as follows
\begin{equation}\label{hyper expand appendix B}
{}_2{F_1}\left( {a,b;c;z} \right) = \sum\limits_{n = 0}^N {\frac{{{{(a)}_n}{{(b)}_n}}}{{{{(c)}_n}n!}}{z^n}}.
\end{equation}

Therefore, the result in~\eqref{Laplace transform general} can be further transformed into
\begin{equation}\label{expanded laplace transform in appendix B}
\begin{aligned}
&{{\mathcal{L}}_{g,i}}\left( { {{\rho _i}} } \right) = \prod\limits_{c = i + 1}^M {} \frac{1}{{(2c - 1)}}\sum\limits_{n = 0}^N {\frac{{{{(1)}_n}{{( - \delta_g )}_n}}}{{{{(1 - \delta_g )}_n}n!}} }\\
& \times {{\left( { - {\tau _i}{{\left( {\frac{M}{R}} \right)}^{\alpha_g} }} \right)}^n}   {\left( {{c^{ - n{\alpha_g}  + 2}} - {{(c - 1)}^{ - n{\alpha_g}  + 2}}} \right)} {r^{n{\alpha_g} }}.
\end{aligned}
\end{equation}

By substituting~\eqref{expanded laplace transform in appendix B} into~\eqref{theorem 1 i>1}, one can obtain
\begin{equation}\label{last expression in appendix B}
{{\rm{P}}_i}({\tau _i}) = {\phi _1}\int_{\frac{{(i - 1)R}}{M}}^{\frac{{iR}}{M}} {} {r^{n{\alpha_g}  + 1}}{e^{ - {\phi _2}{r^{\alpha_g} }}}dr.
\end{equation}
By using $t = {\phi _2}{r^{\alpha_g} }$, and by applying~\cite[eq. (8.350.1)]{Table_of_integrals}, the coverage probability in~\eqref{corolary low tareget rate expression} can be obtained. The proof is complete.

\numberwithin{equation}{section}
\section*{Appendix~D: Proof of Theorem~\ref{theorem aerial coverage probability 2users gamma}} \label{Appendix:Ds}
\renewcommand{\theequation}{D.\arabic{equation}}
\setcounter{equation}{0}

We first derive the conditional coverage probability as follows
\begin{equation}\label{interference limited appendix D}
\begin{aligned}
&{{\rm{P}}_{u,1}}({\tau _1}) = {\mathbb{ E}_I}\left\{ {SINR > {\tau _1}} \right\} \\
& \mathop  = \limits^{(a)} {\mathbb{ E}_I}\left\{ {\frac{{\gamma (m,m{\rho _i}{r^\alpha }{I_u})}}{{\Gamma (m)}}} \right\}\\
& \mathop  = \limits^{(b)}  {\mathbb{ E}_I}\left\{ {\sum\limits_{k = 0}^{m - 1} {\frac{{{{( - m{\rho _i}{r^\alpha }I_u)}^k}}}{{k!}}\exp \left( {m\ln \left( I_u \right)} \right)} } \right\},
\end{aligned}
\end{equation}
where ${I_u} = {I_{u,i}} + {I_g}$, (a) follows the cumulative distribution function of the Gamma random variable, and (b) is obtained from the definition of incomplete Gamma function when $m$ is an integer.

For simplicity, we evaluate the coverage probability of interference limited scenario in the two-device case. Then, by applying Fa \`{a} di Bruno's formula, the derivation can be further transformed into
\begin{equation}\label{interference limited appendix D faadi}
\begin{aligned}
&{{\rm{P}}_{u,1}}({\tau _1}) \\
& ={\mathbb{ E}_I}\left\{ {\sum\limits_{k = 0}^{m - 1} {\frac{{{{( - 1)}^k}}}{{k!}}{{\left[ {\frac{{{\partial ^k}\left( {{{\left( {{\mathcal{L}_{u,i}}(s) + {\mathcal{L}_g}(s)} \right)}^m}} \right)}}{{\partial {s^k}}}} \right]}_{s = {\rho _i}{r^\alpha }}}} } \right\}\\
&= \int\limits_{{r_o}}^{\frac{R}{2}} {} {f_{u,1}}(r){\sum\limits_{k = 0}^{m - 1} {\frac{{{{( - 1)}^k}}}{{k!}}\left[ {\frac{{{\partial ^k}\left( {{\left( {{\mathcal{L}_{u,i}}(s) + {\mathcal{L}_g}(s)} \right)}^m} \right)}}{{\partial {s^k}}}} \right]} _{s = {\rho _i}{r^\alpha }}}dr.
\end{aligned}
\end{equation}

Thus, the overall coverage probability can be obtained by substituting the Laplace transform of interference distribution in~{\textbf {Lemma~\ref{lemma5: aerial device general expression interference for i-th IoT}}} and~{\textbf {Lemma~\ref{lemma6:general expression interference for i-th IoT}}.

\bibliographystyle{IEEEtran}
\bibliography{IEEEabrv,bib2018}

\end{document}